# Sensors in Distributed Mixed Reality Environments

Felix G. HAMZA-LUP
Computer Science, Armstrong Atlantic State University
Savannah, GA 31419

and

Charles HUGHES[1], Jannick P. ROLLAND[1,2]
[1]School of Electrical Eng. and Computer Science
[2]School of Optics-CREOL
University of Central Florida
Orlando, FL 32816

**ABSTRACT**

With the advances in sensors and computer networks an increased number of Mixed Reality (MR) applications require large amounts of information from the real world. Such information is collected through sensors (e.g. position and orientation tracking sensors). These sensors collect data from the physical environment in real-time at different locations and a distributed system connecting them must assure data distribution among collaborative sites at interactive speeds.

We propose a new architecture for sensor based interactive distributed environments that falls in-between the atomistic peer-to-peer model and the traditional client-server model. Each node in the system is autonomous and fully manages its resources and connectivity. The dynamic behavior of the nodes is triggered by the human participants that manipulate the sensors attached to the nodes.

## 1. INTRODUCTION

In the majority of network collaborative Mixed Reality (MR) environments information is collected from participants through graphical user interfaces. In a limited number of cases, head tracking information is embedded in a collaborative MR session. Even in these cases the data is used only locally. Advances in sensors and computer networks have triggered an increase in the number of MR applications that entail large amounts of information from external devices (e.g. tracking systems). Sensor data is especially required to render the position and orientation of the virtual components of the scene in MR applications where a smooth blending of the real and virtual is desired.

We propose a new architecture for sensor based interactive distributed MR environments that falls in-between the atomistic peer-to-peer model and the traditional client-server model. The architecture facilitates the development of distributed MR collaborative applications in which data collected from real-time sensors is shared among the users. Each node in the system is autonomous and fully manages its resources and connectivity. The dynamic behavior of the nodes is dictated by the human participants who manipulate the sensors attached to those nodes.

The paper is structured as follows. Section 2 reviews related work. Section 3 presents the main paradigms used throughout the paper and presents a formal categorization of distributed MR application participants and system nodes. Section 4 gives an overview of the proposed architecture representing a distributed system node as a state machine. In Section 5 the behavior agent is detailed together with its load balancing capability, followed by a discussion regarding MR application Quality of Service requirements. Section 6 presents an application example built on the proposed architecture, followed by concluding remarks in Section 7.

## 2. RELATED WORK

One of the first and most intensive efforts in building a networked simulation was the SIMNET project started in 1983 [1] followed a few years later by the Naval Postgraduate School's NPSNet [2]. The software architecture in both cases was event driven, and a set of predictive modeling algorithms (e.g., dead reckoning) were embedded to compensate for the network delays and to allow the system to scale beyond a local area network (LAN). Important contributions of these systems included the protocol data unit (PDU) that allowed the distribution of simulation data among participants and dead-reckoning algorithms that ensured a reasonable maintenance of dynamic shared state [6]. Later the Distributed Interactive Simulation (DIS) project improved the PDU and this led to the emergence of the IEEE 1278 Standards for Distributed Interactive Simulation and its follow-on IEEE P1516 (see [3] for a discussion of these efforts). A large number of other systems have been created to distribute interactive graphical applications over a set of distributed nodes (e.g. Atlas [4], Paradise [5], DEVA3 VR [7], and MASSIVE-3 [8]).

From the perspective of the results reported here, research in distributed VR can be divided into four categories [9]. The first consists of approaches to optimize the communication protocol through packet compression and packet aggregation [10]. The second is focused on reducing the bandwidth throughout the system, and includes data visibility management, which makes use of area of interest (AOI) management [8] and multicasting [11]. Taking advantage of the human perceptual limitations, like visual and temporal perception [12], constitutes the third category. The fourth category deals with the system architecture. Most distributed virtual environment systems are built on the traditional client-server architecture and are composed of nodes with static behavior, i.e., the server node functions as a data distributor (forwarder) while the client nodes act as data producers and/or consumers. Once the distributed application is deployed, the nodes cannot change their core functionality and data distribution is often bottlenecked by the sever capacity.

Several research efforts have concentrated on the development of sophisticated middleware on top of client-server models for distributed data sharing through remote method calls using an object oriented approach (e.g. Repo-3D [13], Avocado [14]). A drawback of building the middleware frameworks on the distributed object model is the additional delays caused by the software layers, making it difficult to maintain an interactive behavior. A few research efforts [15] have concentrated on the fundamental system architecture attempting to define a spectrum of communication architectures. While each point in the spectrum offers its own performance characteristics the participating nodes cannot change their core functionality based on the user's interactions.

With advances in computer graphics and tracking systems the research community has shifted attention to collaborative environments that span the entire virtuality continuum [16], i.e. Mixed Reality and a subset of MR, Augmented Reality (AR) [17] [18]. AR systems were proposed in the mid '90s as tools to assist different fields: medicine [19], complex assembly labeling [20] and construction labeling [21]. A significant leap towards entertainment based MR systems and applications, was provided by the MR Project [22] developed in Japan and presented at ISMR'99. Projects with similar goals can be found in [23] [24]. These projects brought to life practical collaborative applications for different domains based on MR paradigms. As a common feature, the applications emerging from these projects make extensive use of sensors (tracking, haptic etc) although they are based on a local collaboration paradigm and are built on the traditional client-server architecture. None of the research efforts concurrent to these projects consider the possibility of collecting and distributing real-time sensory information from multiple remote sensors in the collaborative environment.

Considering the preceding systems design's advantages and disadvantages, we introduce a novel architecture in which nodes have a dynamic behavior dictated by the participants attached to those nodes. A prototype application following the proposed architecture has been implemented as both a proof of concept and as a way of investigating the efficiency of such a distributed system in terms of capturing and sharing the sensory data at interactive speeds.

### 3. SENSORS, NODES AND PARTICIPANTS

Collaborative MR applications usually involve the interaction of several participants. While some of the participants actively modify the shared scene, other participants are passive, in the sense that they do not interact with the shared scene. From this point of view we define two categories of participants: ***active participants*** and ***passive participants***. An active participant triggers modifications of the virtual components of the scene from a graphical interface or through a sensor attached to his/her node. Passive participants do not trigger any modifications of the shared scene. They receive visual, haptic and/or audio feedback from the environment. The active/passive attributes of the participants can dynamically change during collaboration. An active participant can become passive and vice versa depending of the collaboration needs.

In what follows, let us define ***a node*** in the distributed system as a computing device that allows a participant to interact with the MR environment. In a distributed collaborative environment, participants need to exchange a wide variety of data (e.g. position/orientation data, virtual components attributes etc.). With the advances in sensor technology, we envision that in future systems a significant amount of data will be collected from sensors and devices attached to the participating nodes. Without loss of generality, we will consider sensors that provide position and orientation information and peripheral devices (e.g. mouse, keyboard) that allow interaction with the virtual components of the scene. The discussion can easily be extended to other types of sensors (e.g. haptic) and other devices that can be part of the distributed system's resources.

Sensors are interaction tools from the point of view of the participant. ***Sensors*** (e.g. attached to a glove like device) provide real-time information about the position and orientation of the real objects (e.g. user's hand). This information may be used by the distributed system's nodes to render the virtual components properly registered (in 3D space) with the real components. An interactive collaborative environment must ensure that the data captured by sensors is distributed with minimum delay to all participants. Moreover each node in the system will need to exchange its sensory data with all or a predefined subset of the system's nodes. A pure centralized data distribution approach (e.g. client-server) would not be efficient because of the additional delay associated with the data collection stage followed by the data distribution. An atomistic peer-to-peer approach would not fit either because of the additional overhead in data distribution. Each node would have to exchange data with all the other nodes. As a fundamental property, the nodes in a sensor based MR system may act as data producers, consumers and distributors, simultaneously.

Before we introduce the details of the hybrid nodes architecture let us define four types or running modes for the distributed system's nodes: active nodes, passive nodes, active forward nodes and passive forward nodes:

- *active nodes* - An active node represents an active participant in the MR environment. Each active node collects data from its sensors and is responsible for making the data available to interested peers as quickly as possible.
- *passive nodes* - A passive node does not inject any information into the MR environment. The node uses the information provided by active and forward nodes to render the shared scene.
- *active forward node* - As the name indicates, a forward node forwards data. The forward node can be *active* or *passive*. An active forward node injects its own as well as the forwarded data into the system.
- *passive forward node* - A passive forward node does not inject new information in the system. These nodes act as pure data forwarders.

### 4. ARCHITECTURE OVERVIEW

Let us introduce a novel architecture and discuss its suitability for the development of sensor based distributed MR applications. The design falls in-between the atomistic peer-to-peer and client-server model. Each node is autonomous and fully manages its resources and connectivity through a set of software agents: a GUI agent, a sensor agent, a rendering agent, and a behavior agent. These agents run on each node and trigger the node behavior, i.e. a node can switch among any of the four modes described in the previous section.

The *GUI agent* is responsible for displaying the interface and collecting the information given by the participant through the GUI. The *sensor agent* is responsible for collecting the information given by the participant through the sensor(s) attached to the node (e.g. tracking sensor, haptic sensor etc).

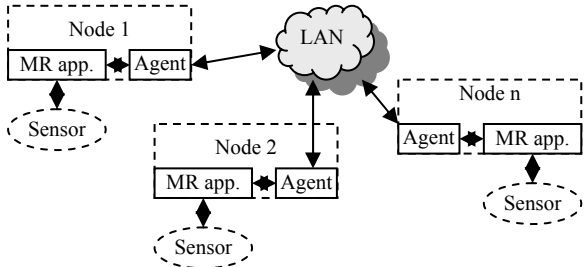

Fig.1. MR distributed environment deployed on a LAN

The *rendering agent* is responsible for rendering the shared scene on the output device (e.g. CRT, LCD, head mounted display etc.) using the data collected from the participating nodes. The *behavior agent* handles new incoming data requests for the active nodes and allows the node to switch among the four modes.

When a participant interacts with the shared scene, the associated node becomes active. We consider the most common interaction scenarios in MR applications: the participant uses a graphical user interface (GUI) for interaction or the participant interacts through a sensor. In the first case, the GUI agent becomes active and makes the data available to the behavior agent while in the second case the sensor agent pulls data from the sensor(s) attached to the node, converts the data into an appropriate format and makes it available to the behavior agent. If requested by other participants, the behavior agent will spawn a server thread making the data available to other nodes hence propagating the local modifications to the shared scene. Otherwise the participant's interaction will affect only the local copy of the shared scene. Regardless of the node mode, the rendering agent is active all the time, given that the scene has to be continuously rendered.

A key characteristic is the architecture capability to map the participant behavior to the distributed systems nodes. The behavior agent is activated by the participant interaction; hence it maps the participant behavior onto the participant's associated node. In other words, if the participant is active, the associated node becomes active and ready to distribute interaction data. If the participant is passive, the associated node becomes passive. Fig.2 illustrates the various agents' states on a passive as well as on an active node.

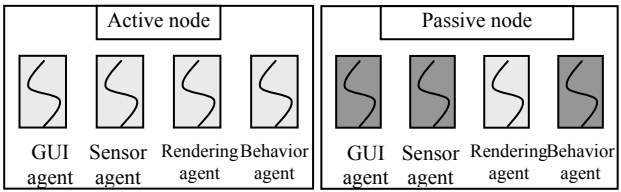

Fig.2 Active, Passive nodes (dark shaded represents inactive agent, light shaded represents active agent)

When the participant does not interact with the shared scene, the GUI agent and the sensor agent are deactivated and the node becomes passive. The behavior agent triggers the forward state of a node. To clarify the possible states (i.e. modes) of a node and the possible transitions, in the next section we describe a state machine that approximates the hybrid node concept.

**Hybrid Nodes as State Machines**
Let's denote the states of a node as: {A, P, AF, PF}. "A" stands for "active", "P" for "passive", "AF" for "Active Forwarder" and "PF" for "Passive Forwarder". Let's denote the conditions that trigger the change in state using a binary representation {00, 01, 10, and 11}

- 00 - means that the Sensor agent goes from **inactive** to **active**
- 01 - means that the Sensor agent goes from **active** to **inactive**
- 10 - means that an activation message has been sent for forwarding **on**
- 11 - means that an activation message has been sent for forwarding **off**

Twelve transitions may occur. Table 1 summarizes the possible transitions, the triggering conditions, and the main operations that are executed at the node.

Table 1. Possible transitions for a hybrid node

| Current State | Next State | Condition(s) | Action on the Server Thread |
|---|---|---|---|
| A | P | 01 | Turn off |
| A | AF | 10 | - |
| A | PF | 01 followed by 10 | - |
| P | A | 00 | Turn on |
| P | AF | 00 followed by 10 | Turn on |
| P | PF | 10 | Turn on |
| AF | A | 11 | - |
| AF | P | 01 followed by 11 | Turn off |
| AF | PF | 01 | - |
| PF | A | 00 followed by 11 | - |
| PF | P | 11 | Turn off |
| PF | AF | 00 | - |

The behavior of a node can be represented as a state machine as illustrated in Figure 3.

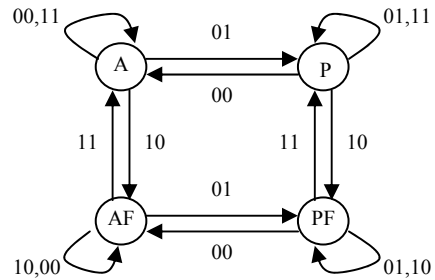

Fig.3 State machine representing the hybrid node behavior

In what follows, we will focus our attention on the component that encapsulates the key characteristic of the proposed design, the behavior agent.

## 5. THE BEHAVIOR AGENT

Our design gravitates around the idea that every node has equal responsibilities. There is no central administration or connection arbiter. Each node renders its local scene based on the data that it receives from its peer nodes. On the other hand each node, while in active or active forward mode, acts as a server multicasting data to the interested nodes. Therefore the design falls in between the atomistic peer-to-peer model and the traditional client-server model.

The behavior agent spawns a thread that controls the activation of the server component of the node. A passive node runs in client mode "consuming" incoming data from the participating nodes. When a passive node becomes active, it means that it acts as a data producer and distributor for the local node and for the other nodes interested in the data.

The atomistic peer-to-peer model has a fundamental bootstrap problem: how to join. Without a central server there is no easy way of determining resource availability in advance. To solve the peer-discovery situation, a new node broadcasts a query to join and awaits response. A complementary problem is: how does a node register itself to receive data from a peer node? It is the responsibility of the behavior agent to advertise that the node has available data from a particular sensor by a short broadcast to all the nodes. If a new peer is interested in the data it will be added by the behavior agent to the node's consumers list. In the next subsection we detail further the producers and consumers list mechanism.

**Consumers and Producers, Load Balancing**
Each node keeps two lists, producers and consumers as shown in Fig.4. The *producers list* contains the node's data providers. The *consumers list* contains the node's current consumers and their connection attributes. An active forwarder adds itself to its *producers list* because it inserts data in the environment.

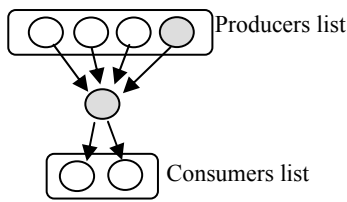

Fig.4 Producers/Consumers list for each node

When a new node requests data from an active node, the behavior agent checks if a direct connection could be established between the two nodes (e.g. application level quality of service QoS considerations might be used). If a direct connection can be established, a client-server relationship is established between the two nodes and the simulation continues. If a direct connection can not be established, the load balancing component of the behavior agent tries to find another potential forwarder node for the data by searching the node's consumers list. If a candidate is found, it becomes the data forwarder (i.e. a forward node). The behavior agent on this node starts a server thread, if it does not have one already, and adds the new client to its consumers list. At the same time, the current node updates its *producers list*. On the other hand, if a candidate is not found, the node can not handle the request at this time. It will reject the request and will invite the new node to join the collaborative MR environment at a later time.

Another approach in finding an appropriate forwarder would be to let the search execute in a recursive manner when a node gets a request that it cannot handle. The node will delegate the request to one of its consumers. If none of the consumers can handle it, they will delegate the request to the next lower level and so on. A depth-first search approach would be more efficient in finding a server for a newly arrived node; however this might lead to a path of considerable depth. Each level on this path would imply additional delays leading to an unacceptable behavior since the system has to respond to the participant's action at interactive speeds. Therefore we choose to search only among the node's direct consumers limiting the length of the virtual path from a data producer to a data consumer to at most two.

**Discussion on Quality of Service for MR/VR Applications**
Allocating sufficient resources to different applications in order to satisfy various requirements is a fundamental problem in distributed systems. Mixed Reality applications deployed on a distributed system infrastructure call for specific quality requirements. With the evolution of wireless networks we experience an increased interest in applications [25] deployed on a wireless infrastructure. Quality of Service is an emerging research area that has developed several approaches to manage application specific quality requirements. These approaches can be grouped in three categories: network QoS [26], utility based QoS [27] and QoS brokers [28]. The Network QoS deals with the management of network resources to provide the QoS guarantees in WANs. In utility based approaches, finite system resources are allocated to multiple QoS dimensions to maximize the overall utility of the application. In the third group, QoS brokers trade the system resources globally and locally to satisfy the application QoS requirements.

An in-depth discussion of the QoS parameters is beyond the scope of this paper. Our goal here is merely to emphasize the architecture's flexibility with regard to the application level QoS parameters. These parameters are considered by the behavior agent for each node when it joins the specific distributed MR application. To maintain a consistent view of the shared scene over all the participants, scene synchronization approaches [29] can be employed up to a certain threshold. Beyond this threshold, application specific QoS requirements are considered.

One of the parameters that has an important effect on distributed applications is the Average Round Trip Time (ARTT) between two participating nodes. Each participating node contains a round trip time table that is frequently updated. This table contains the ARTTs between the node and each of its data consumers. Another parameter that we will consider in future versions is the computational load on each node. The computational load on each node is a factor of the number of sensors it has to manage, the number of producers and consumers it handles, and the rendering complexity of the virtual components in the scene. Each node can measure its average computational load based on the number and complexity of the threads it is executing. Moreover, this takes into consideration the hardware components of the node since the nodes in the system are heterogeneous. The rendering agent is the most sensitive because it directly influences the frame rate. A drop in the frame rate will decrease the quality of the virtual components of the scene and will have negative effects on the user task performance [30].

## 6. APPLICATION

To investigate the efficacy of the hybrid node concept we have developed a prototype application using the Distributed Artificial Reality Environment DARE [31] and VESS [32] frameworks.

**System Setup**
The application was deployed in a distributed system containing five nodes interconnected on a 100Mbps local area network. Additionally two optical tracking sensors were used to insert tracking information into the environment regarding the position and orientation of the virtual objects in the scene.

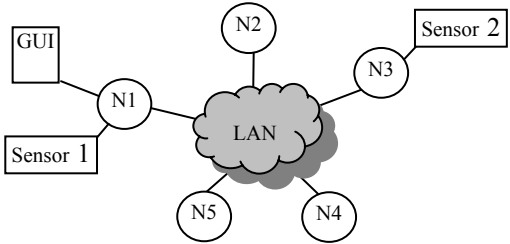

Fig.5 System setup

The participants at nodes N1 and N3, respectively, interact on the shared scene through optical tracking sensors attached to the nodes as shown in Fig. 5. The participant at node N1 can also manipulate particular objects in the shared scene through a graphical interface. At last, participants at nodes N2, N4 and N5 only visualize the virtual shared scene. From the hardware point of view each site consists of one head-mounted display [33], a Linux based PC and an ARC display [34].

**Distributed Interactive 3D Visualization**
At each location, the real environment is augmented with floating 3D objects seen through the head-mounted display as illustrated in Fig.6.

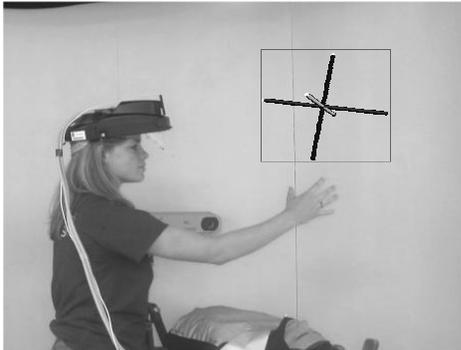

Fig.6 Interactive 3D Visualization participant and virtual 3D cross.

Participants can interact with the 3D models in two ways. Using a graphical user interface they can point in the virtual space to different parts of the virtual objects and they can manipulate them. An alternative interaction method is using a tracking probe attached to an optical sensor (e.g. NDI Polaris Tracking system). By manipulating the probe the user can change the position and orientation of the 3D cross. The application captures the information form the tracking sensors at their update rate (i.e. 60 Hz).

The distributed visualization application implemented on a hybrid node infrastructure is a simple example of a distributed interactive environment that utilizes sensors. We plan to experiment with other interactive distributed mixed reality applications in the near future. One of our goals is to provide a new implementation of the distributed 3D AR training tool for endotracheal intubation [35].

## 7. CONCLUSION

It is clear that distributed collaborative applications involving physically distributed multiple real-time sensors are inherently very complex. It is also clear that interactive distributed MR applications and environments are becoming increasingly common. Thus, the architectural issues in building and organizing the software for such systems must be closely examined.

We have proposed a novel architecture for managing distributed sensors as part of an interactive distributed MR application. The proposed approach allows a dynamic behavior for the distributed systems nodes based on the participants' behavior allowing interactive data capturing and distribution. Furthermore we avoid a complex architecture as we believe simplicity is a key component in developing interactive applications. A subjective assessment of the Distributed Interactive 3D Visualization application interactivity confirmed our expectations. We are in the process of developing a monitoring system that will allow an objective assessment of the proposed architecture performance.

Future work involves refining the architecture and studying its scalability. If the number of remote participants increases we anticipate the need of some centralized control and additional dynamic shared state management techniques to maintain the environment interactivity.

## 8. ACKNOWLEDGEMENTS

We wish to thank our sponsors: The Link Foundation, NSF/ITR IIS-00-820-16, Office of Naval Research Grant N000140310677, and the US Army Simulation, Training, and Instrumentation Command (STRICOM) for their invaluable support for this research.

## 9. REFERENCES

[1] D. Miller and J.A. Thorpe, "**SIMNET: The advent of simulator networking**", Proceedings of IEEE Vol.83, No.8, 1995, pp.1114-1123.

[2] M. Zyda,, D. R. Pratt, J.G. Monahan, and K.P.Wilson, "**NPSNET: Constructing a 3D Virtual World**", Proceedings of ACM Symposium on Interactive 3D Graphics, 1992, pp.147-156.

[3] S. Singhal and M. Zyda, "**The origin of Networked Virtual Environments**", in Networked Virtual Environments: Design and Implementation, Addison Wesley Pub., 1999, pp.19-53.

[4] M. Fairen, and A. Vinacua, "**ATLAS, A Platform for Distributed Graphics Applications**", Proceedings of sixth Eurographics Workshop on Programming Paradigms in Graphics, 1997, pp.91-102.

[5] H.W. Holbrook, S.K. Singhal and D.R. Cheriton, "**Log-Based Receiver-Reliable Multicast for Distributed Interactive Simulation**", in Proceedings of ACM SIGCOMM, 1995, pp. 328-341.

[6] S. Singhal and D.R. Cheriton, "**Exploiting position history for efficient remote rendering in networked virtual reality**", PRESENCE: Teleoperators and Virtual Environments, Vol. 4, No.2, 1995, pp.169-193.

[7] S. Pettifer, J. Cook, J. Marsh, and A. West, "**DEVA3: Architecture for a Large-Scale Distributed Virtual Reality System**", in Proceedings of ACM Virtual Reality Software and Technology (VRST), 2000.

[8] C. Greenhalgh, J. Purbrick, and D. Snowdon, "**Inside MASSIVE3: Flexible Support for Data Consistency and World Structuring**", in Proceedings of ACM Collaborative Virtual Environments (CVE), 2000.

[9] S. Singhal and M. Zyda, "**Resource Management for Scalability and Performance**", In Networked Virtual Environments: Design and Implementation, Addison Wesley Pub., 1999, pp. 181-148.

[10] J.O. Calvin, D.C. Miller, and J. Seeger, "**Application control techniques system architecture**", Technical Report RITN-1001-00, MIT Lincoln Lab., Lexington, MA, 1995.

[11] M. Macedonia, M. Zyda, D. Pratt, P. Donald, P. Brutzman, and P. Barham, "**Exploiting Reality with Multicast Groups**", in IEEE Computer Graphics and Applications, Vol.15, No.5, 1995.

[12] P.M. Sharkey, M.D. Ryan, and D.J. Roberts, "**A local perception filter for distributed virtual environments**", in Proceedings of the IEEE Virtual Reality Annual International Symposium, Atlanta, GA, 1998, pp.242-249.

[13] B. MacIntyre and S. Feiner, "**A Distributed 3D Graphics Library**", in Proceedings of ACM SIGGRAPH, 1998, pp.361-370.

[14] H. Tramberend, "**Avocado: A Distributed Virtual Reality Framework**", in Proceedings of IEEE Virtual Reality, Houston, TX, 1999.

[15] S.K. Singhal, B.Q. Nguyen, R. Redpath, M. Fraenkel, and J. Nguyen, "**InVerse: Designing an interactive universe architecture for scalability and extensibility**", in Proceedings of the Sixth IEEE International Symposium on High-Performance Distributed Computing, IEEE Computer Society, Portland, OR, 1997.

[16] P. Milgram and D. Drascic, "**Perceptual issues in Augmented Reality**", in SPIE Vol. 2653, 1996, pp.123-134.

[17] D. Schmalstieg and G. Hesina, "**Distributed Applications for Collaborative Augmented Reality**", in Proceedings of IEEE Virtual Reality, Orlando, FL, 2002, pp.59-66.

[18] R. Raskar, G. Welch, M. Cutts, A. Lake, L. Stesin and H. Fuchs, "**The Office of the Future : A Unified Approach to Image-Based Modeling and Spatially Immersive Displays**", Proceedings of ACM SIGGRAPH, Orlando FL, 1998.

[19] H. Fuchs, A. State, M. Livingston, W. Garrett, G. Hirota, M. Whitton and E. Pisano, "**Virtual Environments Technology to Aid Needle Biopsies of the Breast: An Example of Real-Time Data Fusion**", in Proceedings of Medicine Meets Virtual Reality, Vol.4, IOS Press, Amsterdam, 1996.

[20] T.P. Caudell and D.W. Mizell, "**Augmented Reality: An Application of Heads-Up Display Technology to Manual Manufacturing Processes**", in Proceedings IEEE Hawaii International Conference on Systems Sciences, 1992, pp.659-669.

[21] A. Webster, S. Feiner, B. MacIntyre, W. Massie and T. Krueger, "**Augmented Reality in Architectural Construction, Inspection and Renovation**", Proceedings of ASCE Third Congress on Computing in Civil Engineering, Anaheim, CA, 1996, pp.913-919.

[22] H. Tamura, "**Overview and final results of the MR project**", in Proceedings of International Symposium on Mixed Reality (ISMR), 2001, pp.97-104.

[23] **ARVIKA** http://www.arvika.de/www/e/home/home.htm

[24] C.B. Stapleton, C. E. Hughes, J. M. Moshell, P. Micikevicius and M. Altman, "**Applying Mixed Reality to Entertainment**", IEEE Computer Vol.35, No.12, 2002, pp. 122-124.

[25] M.A. Livingston, L. Rosenblum, S. Julier, D. Brown, Y. Baillot, J.E. Swan, J.L. Gabbard, D. Hix, "**An Augmented Reality System for Military Operations in Urban Terrain**" in Interservice/Industry Training, Simulation & Education Conference, Orlando, FL, 2002.

[26] R. Braden, L. Zhang, S. Berson, S. Herzog and S. Jamin, "**Resource ReSerVation Protocol (RSVP) -- Version 1 Functional Specification**", in RFC 2205, September 1997.

[27] C. Hoover, J. Hansen, P. Koopman, and S. Tamboli, "**The Amaranth Framework: Probabilistic, Utility-Based Quality of Service Management for High-Assurance Computing**", in 4th IEEE International Symposium on High-Assurance Systems Engineering, 1999.

[28] N. Klara and J.M. Smith, "**The QoS Broker**", in IEEE Multimedia Magazine, Vol.2, No.1, 1995, pp. 53-67.

[29] F. G. Hamza-Lup and J. P. Rolland, "**Adaptive Scene Synchronization for Virtual and Mixed Reality Environments**", in IEEE Virtual Reality 2004, Chicago, IL, March 2004.

[30] B. Watson, V. Spaulding, N. Walker, and Ribarsky W. "**Evaluation of the Effects of Frame Time Variation on VR Task Performance**", in IEEE Virtual Reality Annual Symposium (VRAIS), 1997, pp.38-44.

[31] F. G. Hamza-Lup, L. Davis, J. P. Rolland, and C. Hughes, "**Where Digital meets Physical – Distributed Augmented Reality Environments**", in ACM Crossroads, Vol. 9, No.3. [online], 2003.

[32] J. Daly, B. Kline, and G. Martin, "**VESS: Coordinating Graphics, Audio, and User Interaction in Virtual Reality Applications**", in Proceedings of IEEE Virtual Reality, 2002.

[33] J.P. Rolland, F. Biocca, H. Hua, Y. Ha, C. Gao, and O. Harrisson, "**Teleportal Augmented Reality System: Integrating virtual objects, remote collaborators, and physical reality for distributed networked manufacturing**." [in press] Springer-Verlag, 2003.

[34] L. Davis, J. Rolland, F. Hamza-Lup, Y. Ha, J. Norfleet, B. Pettitt, and C. Imielinska, "**Enabling a Continuum of Virtual Environment Experiences**", in IEEE Computer Graphics and Applications, Vol.23, No.2, 2003, pp.10-12.

[35] J.P. Rolland, L. Davis, and F. Hamza-Lup, "**Development of a training tool for endotracheal intubation: Distributed Augmented Reality**". in Medicine Meets Virtual Reality (MMVR), Vol. 11, 2003, pp. 288-294.